# StudyMe: A New Mobile App for User-Centric N-of-1 Trials

Alexander M. Zenner[1], Erwin Böttinger[1,2]*, Stefan Konigorski[1,2]*

[1]Digital Health Center, Hasso Plattner Institute for Digital Engineering, University of Potsdam, Potsdam, Germany
[2]Hasso Plattner Institute for Digital Health at Mount Sinai, Icahn School of Medicine at Mount Sinai, New York, USA
*Corresponding authors: stefan.konigorski@hpi.de, erwin.boettinger@hpi.de

## Abstract

N-of-1 trials are multi-crossover self-experiments that allow individuals to systematically evaluate the effect of interventions on their personal health goals. Although several tools for N-of-1 trials exist, none support non-experts in conducting their own user-centric trials. In this study we present StudyMe, an open-source mobile application that is freely available from https://play.google.com/store/apps/details?id=health.studyu.me and offers users flexibility and guidance in configuring every component of their trials. We also present research that informed the development of StudyMe. Through an initial survey with 272 participants, we learned that individuals are interested in a variety of personal health aspects and have unique ideas on how to improve them. In an iterative, user-centered development process with intermediate user tests we developed StudyMe that also features an educational part to communicate N-of-1 trial concepts. A final empirical evaluation of StudyMe showed that all participants were able to create their own trials successfully using StudyMe and the app achieved a very good usability rating. Our findings suggest that StudyMe provides a significant step towards enabling individuals to apply a systematic science-oriented approach to personalize health-related interventions and behavior modifications in their everyday lives.

## Author Summary

Information on how to improve personal health is widely available, but it can be difficult to know what works for oneself. One way to find out is performing a scientific experiment called an N-of-1 trial, where a person tries out one or more things in a predefined schedule and collects data to see what helps him or her in reaching a health goal. In the first part of this study, we describe results from a survey including 272 participants, which revealed that individuals are interested in many different health aspects and things to try out. Afterwards, we developed our app StudyMe that we present here. StudyMe is the first app to guide users in creating and running their own fully personalized and customizable N-of-1 trials. An empirical evaluation showed that StudyMe has very good usability and every participant successfully created a unique trial. The flexibility and guidance that StudyMe provides empowers individuals to apply the scientific method of N-of-1 trials to their personal health in everyday live.

# Introduction

Self-experimentation is an intuitive approach whenever the best course of action to improve one's health is unknown (1,2). Yet, it is not available to everyone in a meaningful scientific and easy-to-use way. In clinical settings, N-of-1 trials have become the new gold standard for evaluating interventions on a personal level, when researchers or physicians design a systematic comparison of treatment options for individuals (3,4). In contrast to randomized controlled trials, which provide evidence of what likely works for the average person by evaluating treatments on a population level, N-of-1 trials provide a method to determine what works best for the individual directly. To do so, treatments are applied in a sequence of phases with cross-over.

In this study, we investigate how we can enable individuals without medical expertise to create their own N-of-1 trials. We aim to empower them with a new open-source mobile application, called StudyMe, that guides them in doing so. To give the users full control, we define the concept of *user-centric* N-of-1 trials, meaning trials in which each trial component is configurable by the user. We see four essential components that make up a trial: goal, interventions, measures and schedule. The goal is what the user wants to achieve for his or her health, interventions are one or more behavioral modifications and/or treatments being evaluated, measures are the data collected to evaluate achievement of the goal, and the schedule concerns settings related to the trial's length and phases. We focus on two different designs for N-of-1 trials: evaluating a single intervention by switching between intervention and no-intervention phases (withdrawal design), and comparing two interventions (alternating-treatment design) (5). We have a shared vision of a future in which individuals can and do conduct their own N-of-1 trials (1,2), empowered by StudyMe, to explore ways to improve their health in a scientific and yet convenient way.

Several mobile applications exist that allow individuals to gain insights into their health. General self-tracking applications, such as Apple Health and Google Fit, can be used to gather data on various health aspects. More specific self-tracking applications are geared towards particular symptoms or diseases. Examples include apps such as SleepHealth (6) and mPower (7) that allow users to track factors and symptoms related to sleep and Parkinson's disease, respectively. However, none of these apps provide guidance on how to experimentally evaluate interventions (8). Users are able to collect measures pertinent to their personal health goals in these apps and run their own N-of-1 trials. However, basic requirements to conduct a personal N-of-1 trial, data analysis and interpretation are in general not supported.

More convenient is having an app that guides the user through an N-of-1 trial, by reminding them to follow an intervention or enter a measurement and then visualizes the results. The N1 app allows individuals to compare the effects of two supplements on their cognitive performance (9). SleepCoacher evaluates sleep recommendations from clinicians through short experiments (10). Trialist can compare different interventions and collect measures related to pain and treatment side-effects in trials set up together with a clinician (11). TummyTrials, an app for irritable bowel syndrome-related N-of-1 trials, lets users choose from a list of dietary interventions and select measures and trial duration (12). Besides these apps with specific use cases, there are also N-of-1 trial platforms that can be used for different trials.

QuantifyMe hosts trials created by researchers on a mobile app accessible by users to participate in various experiments with a fixed schedule (13). StudyU, an N-of-1 trial platform that we developed in prior work (8), offers researchers more flexibility regarding the design of their trials, including participant eligibility, measurement collection, and scheduling. However, both these platforms are created to enable researchers, not individuals, to conduct trials and collect data.

In summary, currently available apps generally do not support users with the flexibility and guidance to create and conduct their own user-centric trials in a single app. An advantage of N-of-1 trials is the focus on the individual, yet available tools don't allow users to tailor their trials to their needs and preferences. We believe this is a missed opportunity and aim to give individuals full control to create their own trials through the StudyMe app, without having to rely on access to clinicians and researchers. In the following, we present the StudyMe app and its features, followed by a description of the research we conducted to develop StudyMe. Finally, we present an empirical evaluation of the StudyMe app and a discussion of our methods and findings.

# The StudyMe App

## Overview

StudyMe was designed to enable users to create their own user-centric N-of-1 trials, and then run the created trial. The high-level steps a user completes are shown in Figure 1 and illustrated in detail in Supplementary Video 1.

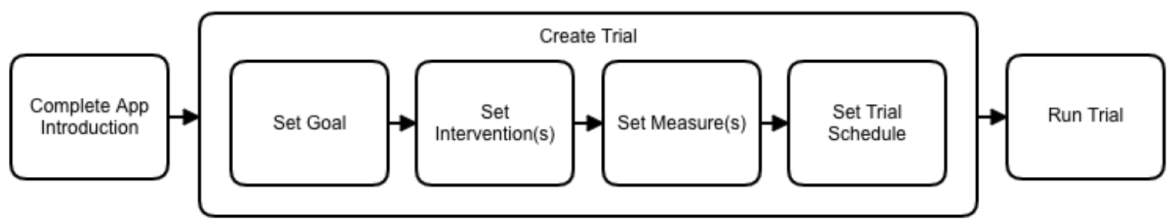

*Figure 1. Overview of the steps the user completes while using StudyMe*

A selection of screenshots, covering the stages of the app, is shown in Figure 2. They show a user's journey, from opening the app, to creating a trial, and running it. After finishing a trial, users can create a completely new trial or one based on their previous trial, allowing for sustained and versatile use of the app.

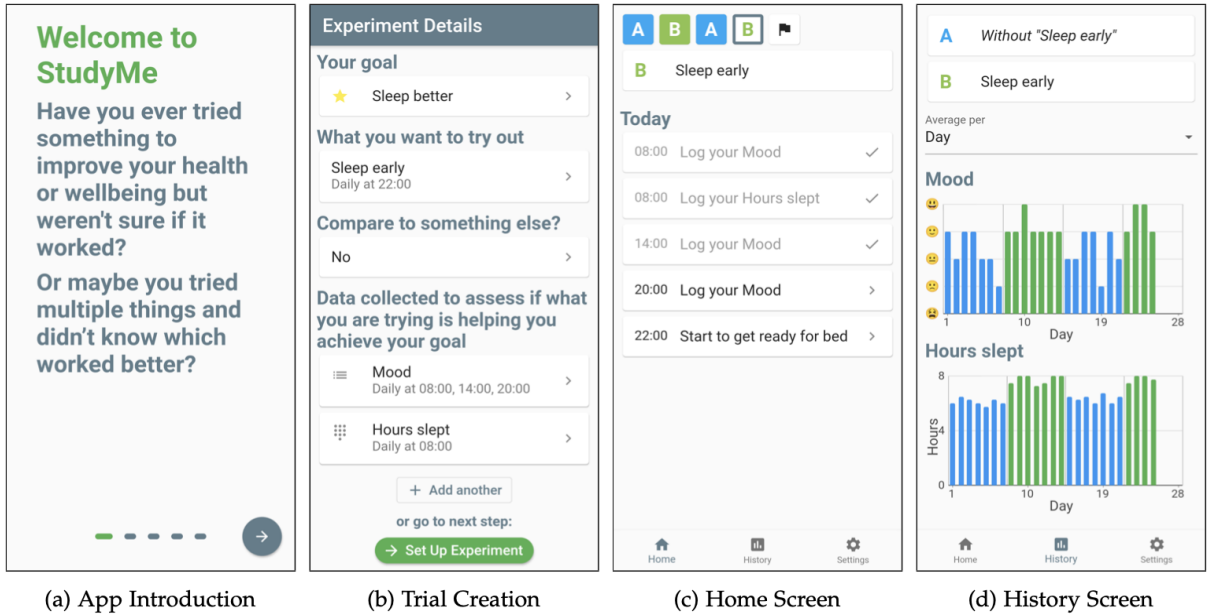

(a) App Introduction  (b) Trial Creation  (c) Home Screen  (d) History Screen

*Figure 2. Selection of screenshots from the StudyMe app.*
*a) Users are first introduced to the app's purpose and the idea of using an experimental methodology. b) Then users create their trials step-by-step, by specifying the different N-of-1 trial components (goal, interventions, measures, schedule). c) Upon starting the trial, the app switches into the "run trial" stage, where trial progress and daily tasks are shown to the users on the Home screen. d) While running the trial, users can view their results on the History screen and compare the data collected during each of the trial phases.*

## Key Features and Design Choices

### Accessible Language and Asking Questions

We reduced N-of-1 trial related jargon in the StudyMe app, substituting technical terms with words we expect to be more commonly understood. A trial is referred to as "your experiment", the outcome as "your goal", interventions as "what / thing you want to try out", and measures as "data you want to collect". Additionally, during the trial creation, the app asks users for what is expected from them. For example, when setting an intervention, the app asks: "What is one thing you want to try out to achieve your goal?"

### Step-by-Step Creation

The centerpiece of trial creation in the StudyMe app is a questionnaire-like screen, called the Experiment Details screen, that guides users step by step through multiple sections, as shown in Figure 3. In Figure 4, we show an example for a sequence of screens when filling a section, in this case, Section 2 regarding interventions ("What you want to try out"). Users are taken to a similar sequence of screens for creating their goal as well as their measures.

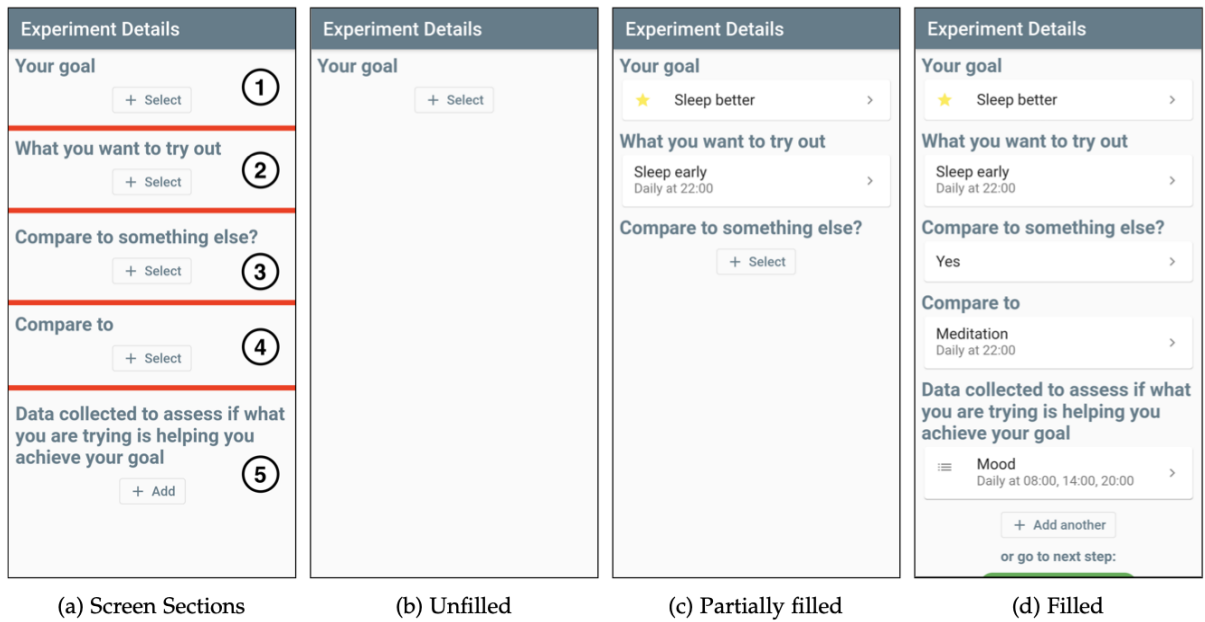

(a) Screen Sections  (b) Unfilled  (c) Partially filled  (d) Filled

*Figure 3. Overview of the Experiment Details screen's sections and how they are filled. a) The screen is divided into five sections that appear one after another. Section 4 appears conditionally with the user's answer to Section 3. b) The screen starts out unfilled. c) Users progress section by section setting the trial's goal, interventions and measures. Section 3 configures the trial's design by choosing whether one intervention is evaluated (withdrawal design) or two interventions are compared (alternating-treatment design). d) Once all sections of the Experiment Details screen are filled, users can continue.*

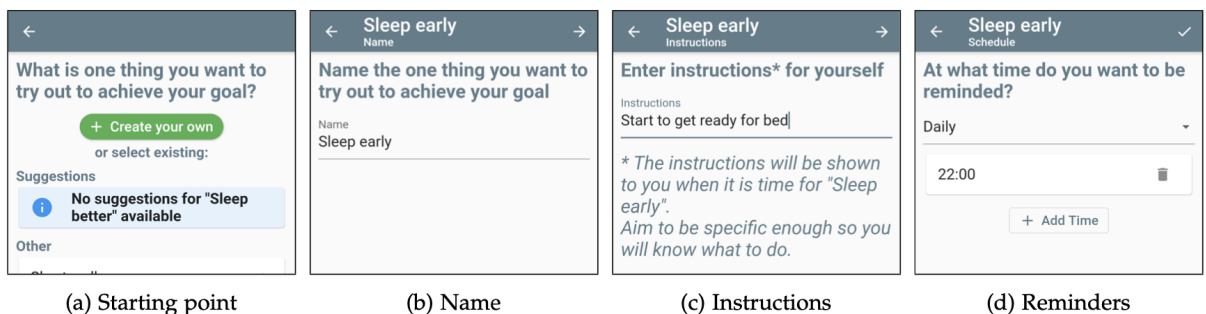

(a) Starting point  (b) Name  (c) Instructions  (d) Reminders

*Figure 4. Sequence of screens that guide users through the creation of their first intervention. a) Users decide if they want to create their own intervention or select an existing intervention. b) If creating their own, they provide a name for the intervention as well as c) instructions for themselves when completing the intervention. d) They are also asked to define when they want to be reminded.*

## Different Trial Designs

During trial creation, users are asked if they would like to compare the first intervention they entered to another one (see Figure 3a). This determines which of the two trial designs will be used. In the withdrawal design, the intervention is evaluated by completing intervention and no-intervention phases. In the alternating-treatment design, a second intervention is entered and the two interventions are compared by completing phases in which either the first or second intervention is applied. In Figure 5 we show this deciding question and how the two possible trial designs are explained to the user on the Your Experiment screen that follows the Experiment Details screen.

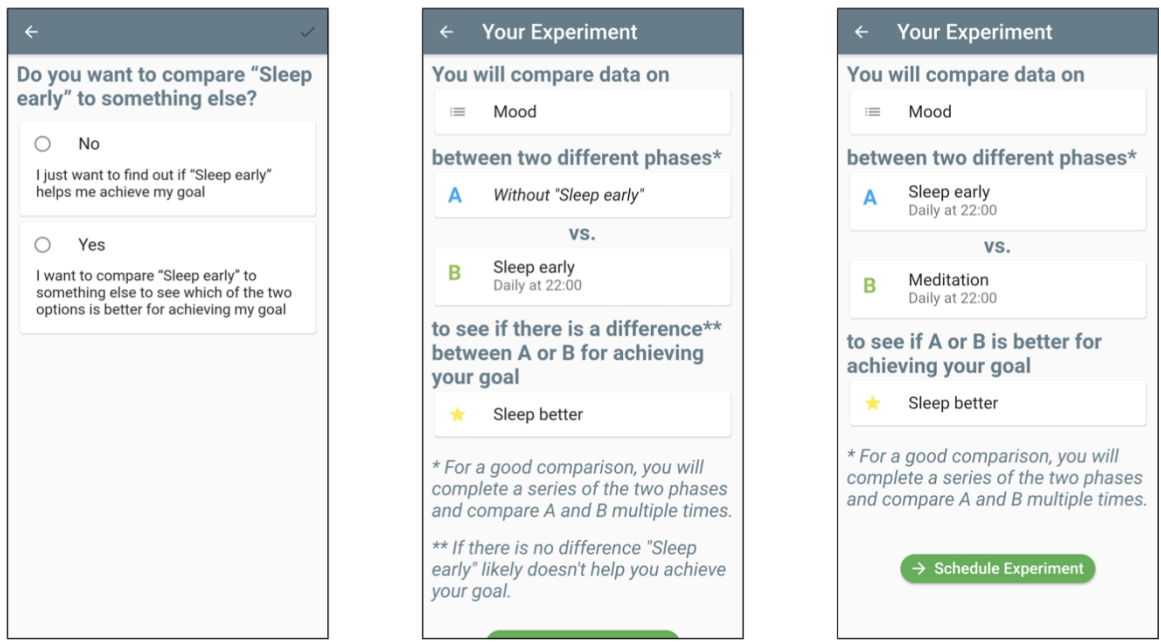

(a) Question used to decide on trial design.

(b) Trial with withdrawal design.

(c) Trial with alternating-treatment design.

*Figure 5. The two trial designs supported by StudyMe.*
*a) The question used to decide between the two designs. The trial is summarized on the Your Experiment screen: b) If the user answered "No", the app creates a trial with a withdrawal design. c) If answered "Yes", two interventions will be compared with an alternating-treatment design.*

## Flexible Scheduling

The last step of creating a trial in StudyMe is defining its schedule. Once the users have filled and reviewed all the other settings, as described in the previous two sections, the app sets a default schedule that the users can edit as shown in Figure 6. The default schedule is set to the phase sequence *ABAB*, with each phase lasting seven days. We've followed recommendations in the literature (14,15) to maintain a balanced phase sequence, as an unbalanced sequence such as *AABB* can yield challenges in the statistical analysis when temporal effects are present. We do so by pairing phases (*AB* or *BA*) and setting the sequence so that the individual either alternates between phases (*ABAB*) or so that the phases are counterbalanced (*ABBA*).

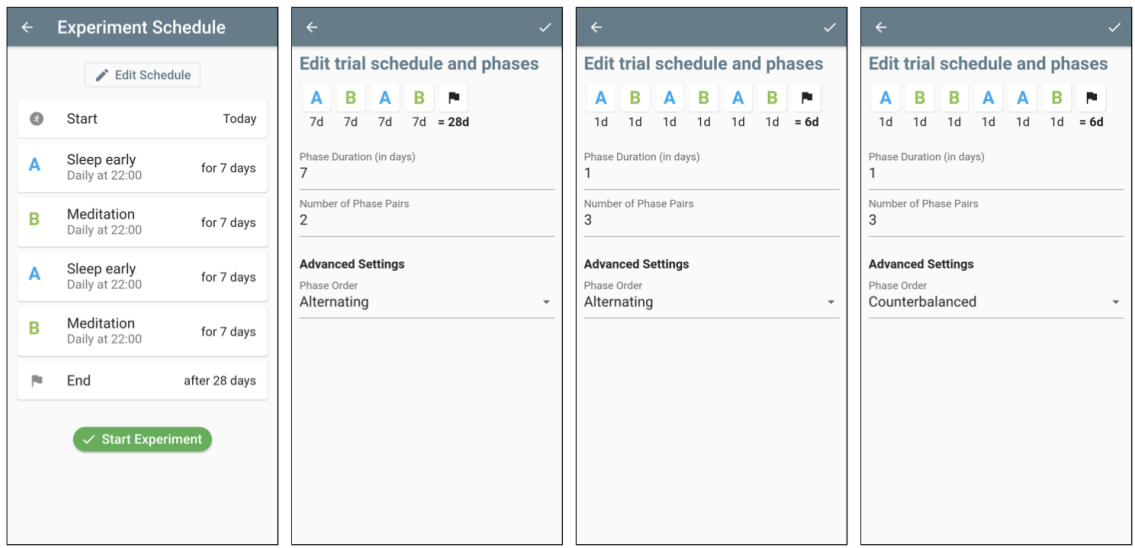

(a) Overview of the schedule  (b) Default schedule settings  (c) More, but shorter, phases  (d) Different phase order

*Figure 6. Different trial schedules in StudyMe.*
*a) An overview of the trial schedule on the Experiment Schedule screen, showing the order and length of the phases as well as the total duration of the trial. b) Selecting 'Edit Schedule' allows the user to change the phase duration, number of phase pairs and choose between two options for determining the phase order. The changes made are reflected dynamically in the minimized overview of the trial's schedule at the top of the screen. c) An example for a trial schedule where the phase duration has been reduced to one day and the number of phase pairs increased to three. d) A schedule where additionally the phase order setting was changed from alternating to counterbalanced.*

## Flexible Reminder Settings

In StudyMe users set reminders for their interventions and measures. These reminders serve the purpose of assisting users in following interventions and taking measures consistently at the same time throughout their trials. As is illustrated in Figure 7, these reminders are set individually and independently for each intervention or measure component, providing high flexibility.

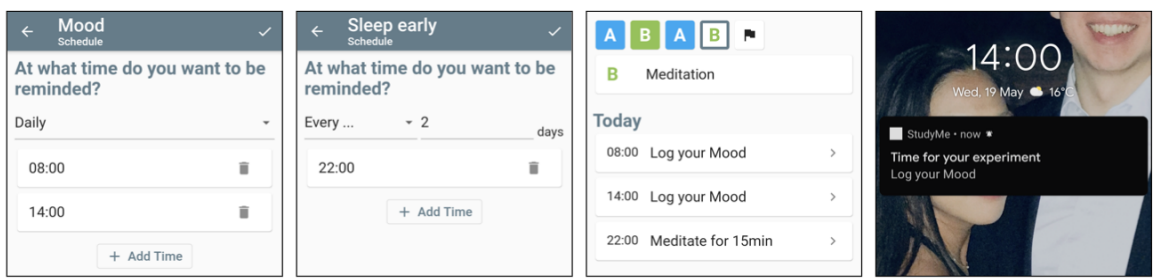

(a) Measure Reminder  (b) Intervention Reminder  (c) Generated Tasks  (d) Notification

*Figure 7. Example for the flexible setting of reminders, their generated tasks and notifications. a) Users can add multiple times in a day and choose if they want to get reminders daily or b) every x days. c) Based on the reminders defined in the trial creation stage, StudyMe generates the tasks to be completed each day and shows them on the home screen. d) Additionally, StudyMe sends notifications to remind users to complete each task at the defined times.*

## Component Libraries

Besides being able to create their own components, users can also select preconfigured components from goal, intervention, and measure libraries. Screenshots of these libraries are shown in Figure 8. Users can gain orientation and inspiration for their own components by looking at the preconfigured ones. Additionally, preconfigured components are quick and easy to add to a trial. They can be treated as templates, as users can further edit them to their liking.

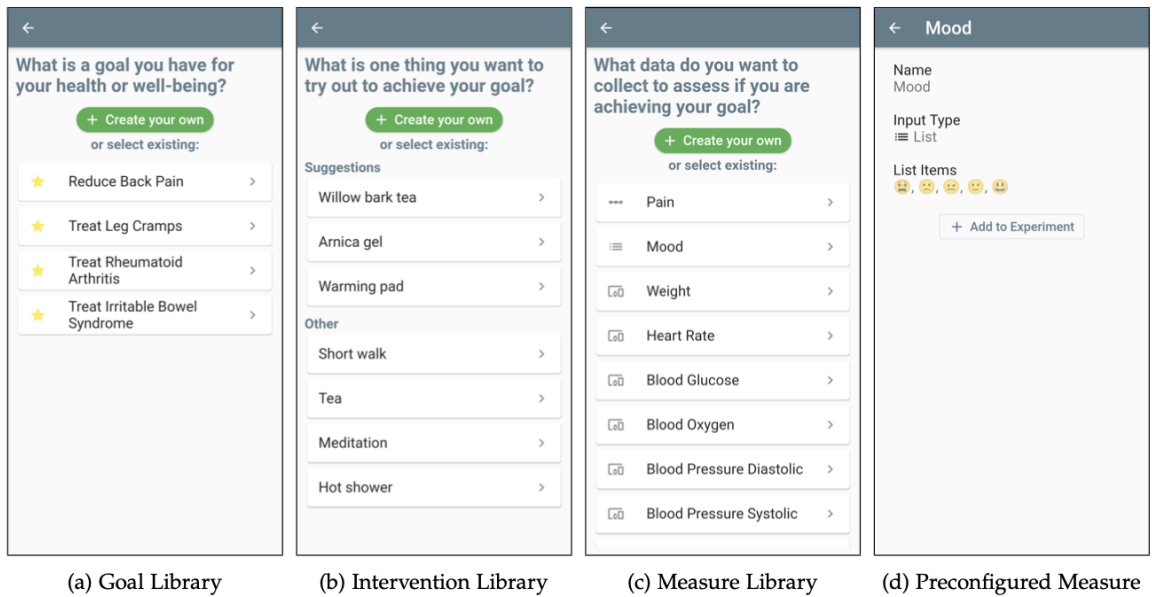

(a) Goal Library  (b) Intervention Library  (c) Measure Library  (d) Preconfigured Measure

Figure 8. The different component libraries.
a) - c) Respective libraries for goal, interventions, and measures. d) Preview screen of a preconfigured measure.

Within the libraries, preconfigured interventions are linked to preconfigured goals, allowing the app to provide specific suggestions together with the other more general examples. The ones currently linked in the app are shown in Table 1.

| Goal | Suggested Interventions |
| --- | --- |
| Reduce back pain | Willow bark tea; Arnica gel; Warning pad |
| Treat leg cramps | Magnesium; Vitamin B12; Massage |
| Treat rheumatoid arthritis | Omega-3 supplement; Olive oil massage; Cold patch |
| Treat irritable bowel syndrome | Gluten-free diet; Fructose-free diet; Low-fibre diet |

Table 1. Linked preconfigured goals and interventions, which are based on example studies developed during prior work on the StudyU platform (8).

## Multiple Measure Input Types

Part of setting up a new measure in StudyMe is to define how the measurements will be inputted into the app. StudyMe provides three input types to offer variety on how measurements can be entered, as shown in Figure 9.

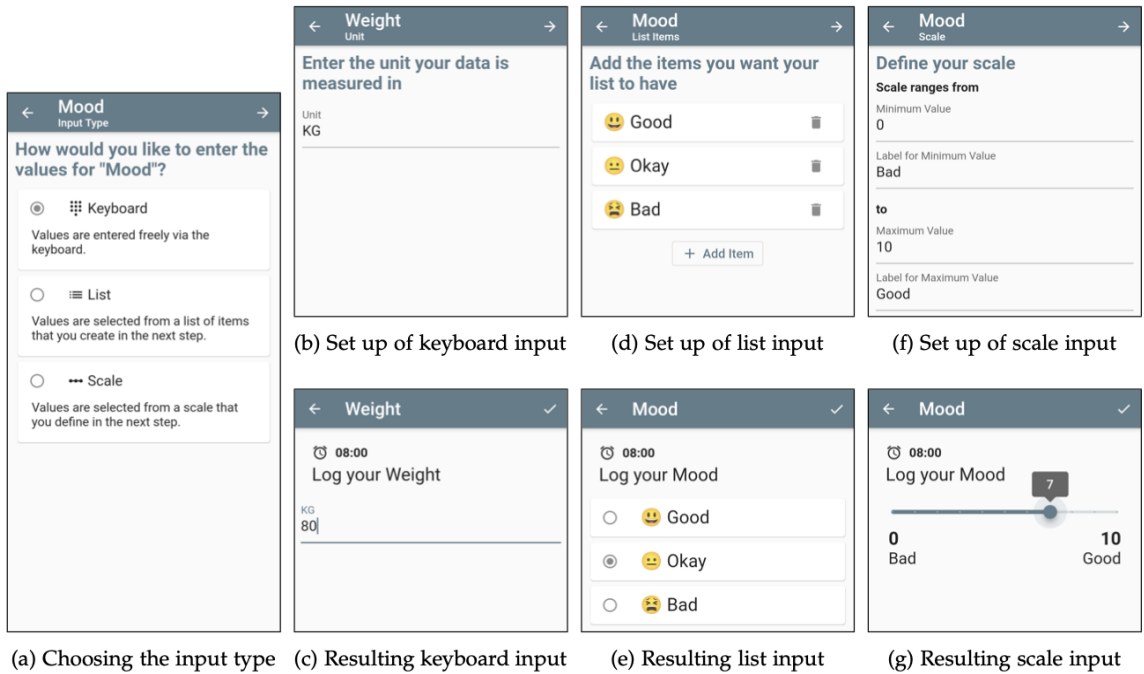

*Figure 9. The different measure input types and how they are set up.*
*a) When creating a measure, users choose the preferred input type. b) The keyboard option allows defining a unit and c) the value is typed in using the phone's keyboard when entering the measurement. d) The list option allows specification of the list items that are offered in e) the resulting list. f) The scale option allows setting the range and annotations that define g) the resulting scale.*

## Implementation & Data Privacy

StudyMe was built with Flutter (https://flutter.dev/), an open-source user interface toolkit written in the Dart programming language. Both Flutter and Dart are developed by Google and can be used to build cross-platform applications, meaning that a single codebase can be compiled to applications for different operating systems. In our case, we focused on developing StudyMe for Android and iOS. The StudyMe Health app is available on the Google Play Store at https://play.google.com/store/apps/details?id=health.studyu.me and can be downloaded for free. No registration is required. The trials and all the other data in StudyMe are stored on the user's device only, to alleviate potential privacy or data security concerns. For an illustration of the study model underlying each created trial, see Supplementary Figure 1. The source code for the app is available on Github at https://github.com/alexanderzenner/studyme.

## Research and Development of StudyMe

We took a user-centered approach of developing StudyMe, with usability as the main focus. StudyMe should help create useful N-of-1 trials for its users and also be straightforward and efficient to understand and interact with. To ensure that individuals that are not familiar with N-of-1 trials can successfully use the app, our process included multiple steps involving research, design, development, and user testing of the application, as shown in

Supplementary Figure 2. The iterations resulted in an adaptation of the language and design to improve the way N-of-1 trial concepts are introduced.

## Survey Methods

In the first stage of the development of StudyMe, we performed an empirical survey in order to get information about which health topics users are interested in and to better understand what trials individuals would want to create as well as what kind of guidance they might need. The survey contained demographic questions on age, sex, and country of residence, and four main questions shown in Table 2. The survey was set up on Google Forms and distributed worldwide over various channels, including institutional email lists, LinkedIn posts, and printed handouts. We aimed for an international sample of survey participants representative of individuals' personal health topics.

| No. | Question (Additional instructions) | Aims to identify |
| --- | --- | --- |
| 1 | Name the one aspect that you want to improve the most about your health or well-being | Health Aspects |
| 2 | Why do you want to improve it? | |
| 3 | In your case, what can you try to improve?<br>(List one or multiple things. Please be as detailed as possible.) | Interventions |
| 4 | How would you assess if this improves the aspect you named?<br>(Please be as detailed as possible.) | Measures |

*Table 2. The four main questions of the survey.*

The qualitative responses were manually coded, assigned to themes in an iterative process based on the concept of an inductive thematic analysis (16), and descriptive statistics were computed.

## Survey Results

272 participants of the survey gave valid responses. 146 participants were female, 125 male, and 1 person identified with 'other'. The mean age was 32 years and the participants were between 16 and 84 years old. We received responses from 22 countries with almost 60% of responses from Germany. Below we present the summarized themes of health aspects, interventions, and measures derived from the participants responses (see Tables 3-5). Responses that weren't codeable due to lack of clarity or misinterpretation of the question asked were excluded.

| Health Aspect Themes | # | % Participants |
|---|---|---|
| Lifestyle | 175 | 64,34 |
| Physical Symptom / Concern | 95 | 34,93 |
| Mental Health | 69 | 25,37 |
| Overall Health / Well-Being | 65 | 23,90 |
| Maintain Health / Prevent Disease | 43 | 15,81 |
| None | 1 | 0,37 |
| Excluded | 8 | 2,94 |

Table 3. Absolute (# Participants) and relative frequencies (% Participants) of responses allocated to health aspect themes derived from responses for questions 1 and 2.
'Lifestyle' includes mentions related to fitness, exercise, weight, diet, productivity, work, stress, and appearance. 'Physical Symptom / Concern' includes mentions related to symptoms, body parts, and diseases. 'Mental Health' includes depression, anxiety, different emotions and feelings, the mind, and cognitive abilities. Some individuals reported wanting to improve their overall health and some wanted to maintain it. The themes are not disjoint and some responses were assigned to multiple themes.

| Intervention Themes | # Participants | % Participants |
|---|---|---|
| Exercise | 153 | 56,25 |
| Diet | 93 | 34,19 |
| Behavior Other | 30 | 11,03 |
| Meditation | 29 | 10,66 |
| Sleep | 26 | 9,56 |
| Thought Other | 26 | 9,56 |
| Time/Work Management | 24 | 8,82 |
| Pharmaceutics | 17 | 6,25 |
| Professional Help | 16 | 5,88 |
| Social | 16 | 5,88 |
| Supplements | 11 | 4,04 |
| Equipment | 9 | 3,31 |
| Other | 24 | 8,82 |
| None | 3 | 1,10 |
| Excluded | 18 | 6,62 |

*Table 4. Absolute and relative frequencies of responses allocated to intervention themes derived from responses for question 3.*

*'Behavior Other' contains mentions related to activities that were not assigned to other themes, for example, improving habits, setting goals, turning off devices, reading, and journaling. 'Thought Other' regards thought-related changes, such as reflecting, thinking happy things, and focusing. 'Social' includes mentions of talking to or meeting friends or family. 'Equipment' includes mentions of objects like "use of a sitting ball" and changes to work setups. The themes are not disjoint and some responses were assigned to multiple themes.*

| Measure Themes | # Participants | % Participants |
|---|---|---|
| Feeling | 89 | 32,72 |
| Physical Performance | 49 | 18,01 |
| Body Composition | 42 | 15,44 |
| Symptoms | 26 | 9,56 |
| Physiological | 20 | 7,35 |
| Productivity | 16 | 5,88 |
| Appearance | 12 | 4,41 |
| Sleep | 10 | 3,68 |
| Stress | 9 | 3,31 |
| Occurence of Sickness | 7 | 2,57 |
| Track Consumption | 6 | 2,21 |
| Other | 44 | 16,18 |
| None | 9 | 3,31 |
| Excluded | 42 | 15,44 |

*Table 5. Absolute and relative frequencies of responses allocated to measure themes derived from responses for question 4.*
*'Track Consumption' includes counting consumption of certain types of meals, frequency of smoking, and tracking calories. Several codes did not fit into one of the other themes and were assigned to 'Other', including quality of relationships, posture, and general well-being. The themes are not disjoint and some responses were assigned to multiple themes.*

One observation in the more detailed analysis of the answers was that some of the larger intervention and measure themes could be further divided into specific sub-themes, while others were unspecific. For example, the intervention theme 'Exercise', contained 114 unspecific mentions such as "do more sports". In comparison, only 55 participants specified the exercise activity and mentioned, e.g., going for a run or stretching. Another observation was that even if participants wanted to improve the same health aspect, they had different ideas for what interventions to try or measures to use. For example, from several individuals that wanted to improve their weight, responses for how they would measure improvement varied, such as "if I lose weight", "more confident when going to the beach", "visually, looking at the mirror", or "fit into my clothes better". Different ways for how individuals would measure progress included Likert scales, counting something, or answering binary questions. We also counted how many responses, for both interventions and measures, mentioned frequencies, such as "check in 3-4 times per day on computer" to measure one's energy levels. 27 intervention responses (9.93%) and 31 measure responses (11.40%) included such frequencies.

## Implications for N-of-1 Trial Apps

With few exceptions, all participants identified a health aspect they wanted to improve as well as one or more interventions they could try out, highlighting the potential relevance of an app for N-of-1 trials. Based on our analysis, what individuals want to improve covers a range of lifestyle-related, physical, and mental health topics. We also identified different themes for interventions and measures. Furthermore, even if participants had stated the same health aspect, what they wanted to try out or measure was different. To allow for this variety, StudyMe has to provide flexibility in what interventions it can be used for as well as what measures individuals can set up. Several of the interventions and measures participants produced were unspecific. Furthermore, only a few participants mentioned frequencies for how often they would do their intervention or take their measurement. StudyMe generates tasks based on what interventions and measures were set. Locke and Latham's goal setting theory suggests that having more specific rather than unspecific task goals leads to better performance (17). Considering this, we focused on providing guidance in StudyMe for creating specific interventions and measures and assisting users in setting reminders for when they want to conduct their interventions and measurements. A possible way to provide guidance is by using examples. Although we received a wide range of responses for each of the questions, certain interventions, such as those related to exercises and diets, and measures, such as those related to subjective feelings, were identified to be most popular. Including examples for those interventions and measures could therefore be helpful.

# Empirical evaluation of StudyMe

The final version of StudyMe was evaluated empirically to (i) see how successful participants were in creating an N-of-1 trial by themselves, (ii) measure the usability of the app, and (iii) obtain further feedback on StudyMe.

## Methods

For the evaluation, we set up a survey on Google Forms with a link for participants to download the app onto their Android device, create a trial, and then answer a set of questions. Participants also reported their created trials, by copying a JavaScript Object Notation (JSON) string from the app and pasting it into the survey, allowing us to analyze the settings they selected. To determine the usability, we used the validated and popular System Usability Scale (SUS) (18) and asked open-ended feedback about the app. Also, a few questions were asked regarding the participants' demographics. The survey was distributed to individuals in our professional network, excluding those that had previously interacted with the app.

# Results

13 participants completed all parts of the evaluation; 1 participant was unable to install the app on her own device and was excluded from the evaluation. The participants had varying knowledge of how medical trials work, were generally interested in personal health topics, and had experience with smartphones and apps. They were aged between 21 and 57 years, nine were men and four were women. 2 responses came from Canada and the other 11 responses from Germany. Figure 10 visualizes all the goal, intervention, and measure components that were created or used by the participants as well as how they were combined in each of the different trials. Four participants created trials that only consisted of preconfigured components chosen from the respective libraries and did not edit the default schedule that StudyMe suggests. None of the trials were completely customized, meaning everyone chose at least one of the preconfigured components and/or did not edit the default schedule. All the created trials were different from each other. Even if participants chose the same preconfigured goal, they still chose different interventions or measures. In all custom-specified trials, the interventions were clearly described, and the reminder features were used and correctly applied.

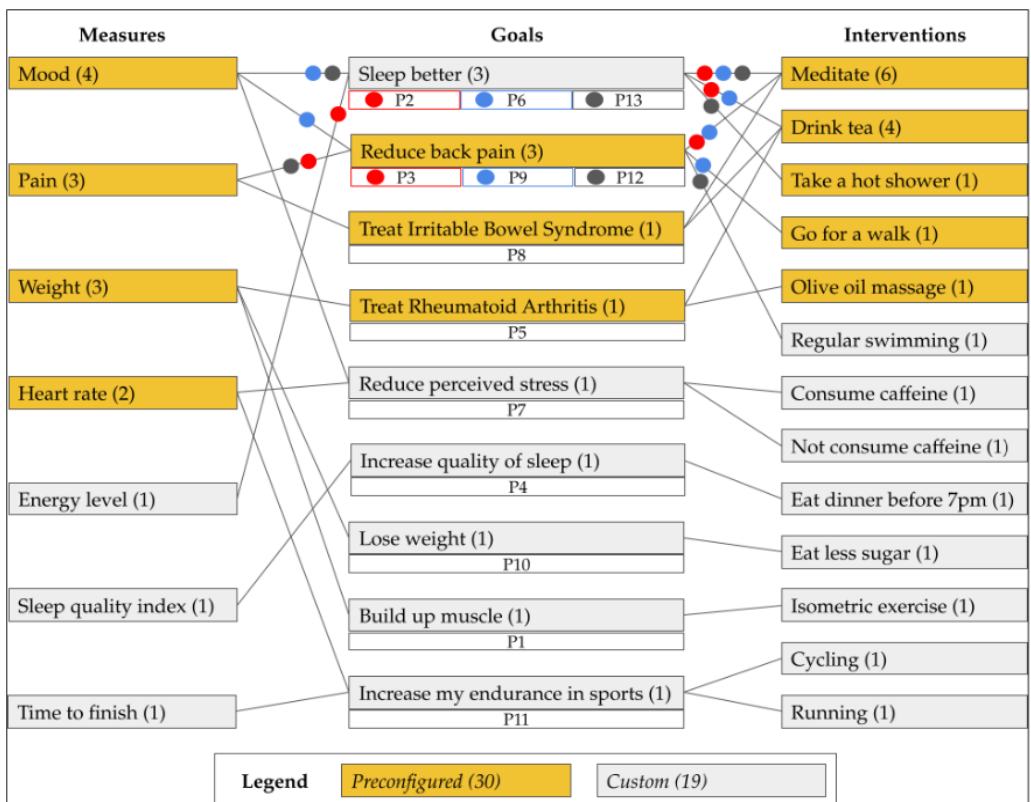

*Figure 10. Visualization of the trials created during the evaluation.*
*The preconfigured and custom measure, goal, and intervention components are shown as rectangles with their frequency in parentheses. Lines between rectangles represent which components were used together. For each goal, we also indicate which participants specified the goal and its connected components (person P1 to P13 below the goal components). In cases where multiple participants had the same goal, the colored circles allow tracing of components that were used by each individual.*

Figure 11 shows the SUS results. The app received a mean SUS score of 82, with 60 as the lowest and 100 as the highest individual SUS score. Overall, the participants mostly agreed with the positive statements and disagreed with the negative statements regarding StudyMe's usability. For more information regarding the created trials, and regarding qualitative feedback on the usability of StudyMe, see the Supplementary Text.

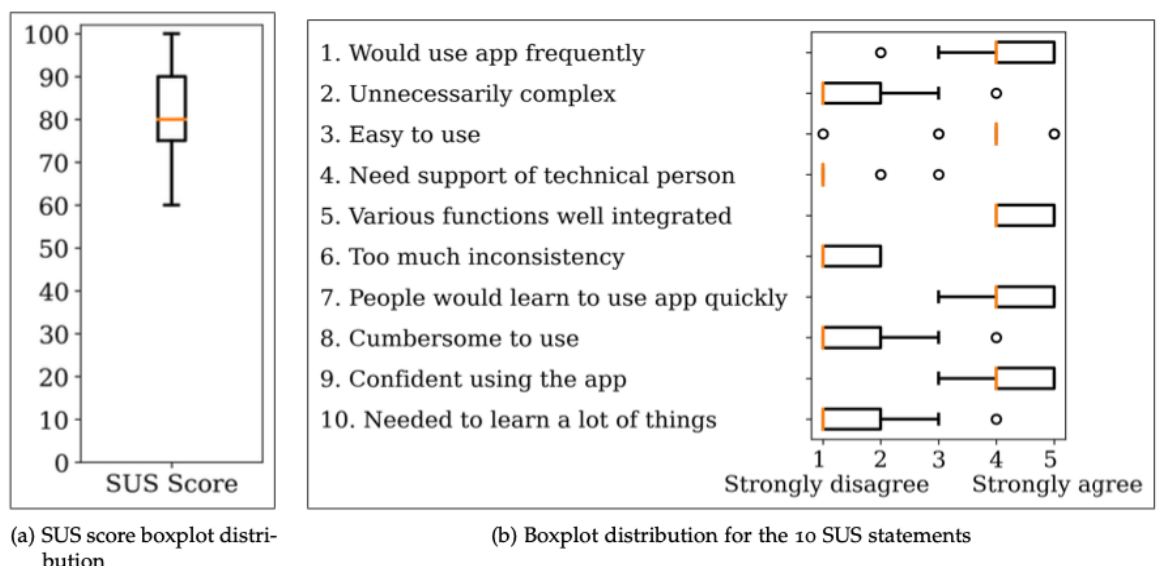

(a) SUS score boxplot distribution

(b) Boxplot distribution for the 10 SUS statements

*Figure 11. System Usability Scale results showing a) the distribution of the SUS scores and b) the distribution of participants' answers for each of the 10 statements.*
*Note: The ten statements were shortened for concise visualization in this figure.*

# Discussion

In this study, we present StudyMe, a new mobile app for user-centric N-of-1 trials. StudyMe provides users with all the features to create a trial, from choosing pre-configured components or creating them freely, to defining reminders and customizing a trial's schedule. The development of StudyMe was based on an empirical survey and multiple iterations of user testing, as well as a final evaluation. Throughout development, we focused on the app's usability, as well as the flexibility and guidance it offers for creating N-of-1 trials. The empirical evaluation yielded a SUS score of 82, which is in the 90-95 percentile range and grants the StudyMe app an A grade (19). The very good usability is further underlined by the qualitative feedback we received and the fact that all participants in the evaluation were able to create their own trials, ranging from improving sleep, to building up muscles, to increasing endurance.

One key aspect of StudyMe and user-centric apps in general is the set-up of the study design of the N-of-1 trial. In tests with a preliminary version of StudyMe, users struggled to specify an intervention schedule on their own. As a result, we restructured the app and included the default schedule that can be edited optionally. In the evaluation of the final version, only two users changed this default schedule. One person increased the length of each phase to 28 days, which is reasonable for evaluating the effects of regular swimming on his back pain, as they are likely to take time to become noticeable. The other person reduced the number of phase pairs to one (phase sequence *AB*), which was not an optimal design for his trial. In future iterations, we plan to provide scheduling suggestions based on automated statistical power calculations using the context of the other trial components.

Other important aspects of any app implementing N-of-1 trials are the suitability of the N-of-1 trial design for the respective study question, the lack of blinding and randomization in StudyMe, as well as potential ethical concerns. Regarding the applicability of StudyMe, the most popular interventions participants mentioned in our survey were exercise and diet-related, which are well-suited for an N-of-1 trial study design (20-22). We opted against limitations on what interventions individuals can use StudyMe for (except for misuse and illegal or unethical studies, which are prohibited by the terms of use), as we aimed to find out what users are interested in. Blinding and randomization are important concepts for reducing the risk of bias in the statistical analysis of trial results (4,14,23). We decided to not include any blinding or randomization functionalities in the current version of StudyMe, as they are challenging to execute in self-defined and self-administered trials, especially if interventions other than drugs are evaluated (24). This is also true for exercises that were frequently mentioned in our survey and that usually require conscious execution. Also, from an individual perspective it might be less relevant to ensure whether an intervention actually has a causal effect or if it is due to a belief that it does, as long as there seemingly is an improvement. Regarding ethical aspects, we believe anything that involves individuals' health should be treated with caution and StudyMe is purposely designed to help individuals in conducting experiments on themselves. For that reason, we only included preconfigured interventions that we expect to be safe, for example, going for a walk. Also, it should be noted that individuals can and do conduct self-experiments with or without being guided by a dedicated mobile application (1). Ultimately, we do not see our app as an alternative to professional health care, but instead envision that individuals use StudyMe after discussing the safety of their trials with their doctors or ideally create the trials together with them.

The reported empirical survey that served as a starting point for creating StudyMe as well as the final empirical evaluation carry some limitations. The sample of our survey on individuals' personal health topics was skewed towards a younger and predominantly German population. Due to the fact that a large part of participants were affiliated with universities, it can also be assumed that we reached a fairly tech-savvy and educated group of individuals. The final empirical evaluation was done with 13 participants, which are likely not representative of a larger population. As such, StudyMe requires further evaluation with users of different backgrounds to ensure its broad applicability.

In future work, we plan to apply StudyMe in long-term studies following individuals running their trials. Furthermore, we envision that the component libraries can be extended to include more preconfigured goals, interventions, and measures. To achieve this, we plan to integrate the StudyMe app with the StudyU platform (8) so that the vetted trials and components created by researchers on StudyU can be offered as inspiration to the users on StudyMe. Beyond that, including features that allow users to share their trials and components with each other would allow individuals to create and run their trials with their friends, family members and a larger community. Part of this could be adding gaming aspects to StudyMe's design, as gamification in mHealth applications has been shown to have a positive effect on individuals' behaviors, especially those involving physical activity (25). We are looking forward to seeing the use of StudyMe in enabling individuals to reach their personal health goals, by testing interventions in a systematic way and contributing to a citizen-empowered transformation of healthcare.

# Acknowledgments

We thank all participants of the survey, intermediate tests, and app evaluation, as well as Stella Leowinata and Babajide Owoyele for their critical feedback and discussion of the app design and development.

# References

1. Neuringer A. Self-experimentation: A call for change. *Behaviorism*. 1981;9(1):79–94.
2. Wolf GI, De Groot M. A conceptual framework for personal science. *Frontiers in Computer Science*. 2020;2(21).
3. Guyatt G, Sackett D, Taylor DW, Ghong J, Roberts R, Pugsley S. Determining optimal therapy — randomized trials in individual patients. *New England Journal of Medicine*. 1986;314(14):889–892.
4. Lillie EO, Patay B, Diamant J, Issell B, Topol EJ, Schork NJ. The n-of-1 clinical trial: The ultimate strategy for individualizing medicine?. *Personalized Medicine*. 2011;8(2):161–173.
5. Tate RL, Perdices M. N-of-1 trials in the behavioral sciences. In: Nikles J, Mitchell G. (eds.) *The Essential Guide to N-of-1 Trials in Health*. Dordrecht: Springer Netherlands; 2015. p. 19–41.
6. Deering S, Pratap A, Suver C, Borelli AJ, Amdur A, Headapohl W, Stepnowsky CJ. Real-world longitudinal data collected from the SleepHealth mobile app study. *Scientific Data*. 2020;7(418).
7. Bot BM, Suver C, Neto EC, Kellen M, Klein A, Bare C, Doerr M, Pratap A, Wilbanks J, Dorsey ER, Friend SH, Trister AD. The mPower study, Parkinson disease mobile data collected using ResearchKit. *Scientific Data*. 2016;3(160011).
8. Konigorski S, Wernicke S, Slosarek T, Zenner AM, Strelow N, Ruether FD, Henschel F, Manaswini M, Pottbäcker F, Edelman JA, Owoyele B, Danieletto M, Golden E, Zweig M, Nadkarni G, Böttinger E. StudyU: A platform for designing and conducting innovative digital N-of-1 trials. *arXiv*. [Preprint] 2020. Available from: https://arxiv.org/abs/2012.14201.
9. Golden E, Johnson M, Jones M, Viglizzo R, Bobe J, Zimmerman N. Measuring the effects of caffeine and L-theanine on cognitive performance: A protocol for self-directed, mobile N-of-1 studies. *Frontiers in Computer Science*. 2020;2(4).
10. Daskalova N, Metaxa-Kakavouli D, Tran A, Nugent N, Boergers J, McGeary J, Huang J. SleepCoacher: A personalized automated self- experimentation system for sleep recommendations. In: *Proceedings of the Annual Symposium on User Interface Software and Technology, Tokyo, Japan*. ACM; 2016, p. 347–358.
11. Barr C, Marois M, Sim I, Schmid CH, Wilsey B, Ward D, Duan N, Hays RD, Selsky J, Servadio J, Schwartz M, Dsouza C, Dhammi N, Holt Z, Baquero V, MacDonald S, Jerant A, Sprinkle R, Kravitz RL. The PREEMPT study - evaluating smartphone-assisted n-of-1 trials in patients with chronic pain: Study protocol for a randomized controlled trial. *Trials*. 2015;16(67).

12. Karkar R, Schroeder J, Epstein DA, Pina LR, Scofield J, Fogarty J, Kientz JA, Munson SA, Vilardaga R, Zia J. TummyTrials: A feasibility study of using self-experimentation to detect individualized food triggers. In: *Proceedings of the CHI Conference on Human Factors in Computing Systems*, *Denver, CO, USA*. ACM; 2017, p. 6850–6863.
13. Taylor S, Sano A, Ferguson C, Mohan A, Picard RW. QuantifyMe: An open-source automated single-case experimental design platform. *Sensors*. 2018;18(4).
14. Kravitz R, Duan N, Eslick I, Gabler NB, Kaplan HC, Larson EB, Pace WD, Schmid CH, Sim I, Vohra S. *Design and Implementation of N-of-1 Trials: A User's Guide*. Rockville, MD, USA: Agency for Healthcare Research and Quality; 2014. Available from: https://effectivehealthcare.ahrq.gov/products/n-1-trials/research-2014-5.
15. Carriere KC, Li Y, Mitchell G, Senior H. Methodological considerations for N-of-1 trials. In: Nikles J, Mitchell G. (eds.) *The Essential Guide to N-of-1 Trials in Health*. Dordrecht: Springer Netherlands; 2015. p. 67–80.
16. Braun V, Clarke C. Using thematic analysis in psychology. *Qualitative Research in Psychology*. 2006;3(2):77–101.
17. Locke EA, Latham GP. *A Theory of Goal Setting & Task Performance*. Englewood Cliffs, NJ, USA: Prentice Hall; 1990.
18. Brooke J. SUS: A 'quick and dirty' usability scale. In: Jordan PW, Thomas B, Weerdmeester BA, McClelland IL. (eds.) *Usability Evaluation in Industry*. London, UK: Taylor & Francis; 1996. p. 189–194.
19. Sauro J, Lewis JR. *Quantifying the User Experience: Practical Statistics for User Research*. Waltham, MA, USA: Elsevier/Morgan Kaufmann; 2012.
20. Yoon S, Schwartz JE, Burg MM, Kronish IM, Alcantara C, Julian J, Parsons F, Davidson KW, Diaz KM. Using Behavioral Analytics to Increase Exercise: A Randomized N-of-1 Study. *Am J Prev Med*. 2018;54(4):559–567.
21. Brannon EE, Cushing CC, Walters RW, Crick C, Noser AE, Mullins LL. Goal feedback from whom? A physical activity intervention using an N-of-1 RCT. *Psychology & Health*. 2017;33(6):701-12.
22. Kaplan HC, Opipari-Arrigan L, Schmid CH, Schuler CL, Saeed S, Braly KL, Burgis JC, Nguyen K, Pilley S, Stone J, Woodward G, Suskind DL. Evaluating the comparative effectiveness of two diets in pediatric inflammatory bowel disease: A study protocol for a series of N-of-1 trials. *Healthcare*. 2019;7(4).
23. Guyatt G, Sackett D, Adachi J, Roberts R, Chong J, Rosenbloom C, Keller J. A clinician's guide for conducting randomized trials in individual patients. *Canadian Medical Association Journal*. 1988;139(6):497–503.
24. Boutron I, Tubach F, Giraudeau B, Ravaud P. Blinding was judged more difficult to achieve and maintain in nonpharmacologic than pharmacologic trials. J*ournal of Clinical Epidemiology*. 2004;57(6):543–550.
25. Johnson D, Deterding S, Kuhn KA, Staneva A, Stoyanov S, Hides L. Gamification for health and wellbeing: A systematic review of the literature. *Internet Interventions*. 2016;6:89–106.

# Supplementary Information

## Supplementary Text: Details on the empirical evaluation of StudyMe

In the following, we provide more details on the results of the empirical evaluation of StudyMe from the user testing with 13 participants.

When analyzing the instructions that participants provided for their custom interventions in the created trials, it was apparent that each of them specified how the intervention would be conducted, for example the person that defined an intervention with the name "cycling" wrote "cycle to university or on the way back home". The participants used the settings of StudyMe to define reminders for their interventions and measures at different times and frequencies. The four participants with sleep-related goals used daily reminders. Three of them (P2, P6, P13) set their intervention reminders to times in the evening between 6:00 p.m. and 10:30 p.m. The fourth person (P4) set the reminders for his intervention "eat dinner before 7pm" to noon. All four set their measure reminders to times in the morning between 6:30 a.m. and noon. Additionally, others set their reminders at different intervals, e.g., P12 who defined to go swimming every three days. 2 of the 13 participants changed the default schedule.

Qualitative feedback was gathered in addition to the SUS score results reported in the main manuscript. When asked what participants liked about the app, many participants mentioned aspects related to the flexibility and guidance that the app offers. They liked that they could create their own experiment. One participant stated that the app "allows me to experiment more than one way to reach my goal" and another that "it's very flexible". With respect to guidance, participants stated that they liked the app's "structured approach", "guided setup", and "well explained process". One participant wrote that it "feels empowering to get a support framework for self-testing". Another liked the "recommendations for studies and actions" as well as the fact that the app uses statements with gaps that she can fill when creating a trial. Other positive remarks regarded "the reminder", "good defaults", and the app's design which was described as "professional-looking" and "extremely clean". It was also explicitly described as user-friendly, and "fast, easy and straight-forward to use". With regard to what could be improved in StudyMe, some participants noted that there were "too many questions in the beginning", that "setting notifications was cumbersome" and that it was "odd having to create separate reminders to collect data". Seemingly contrary to this, one person said that she would actually like "more options to select dates". While the app was designed to reduce the number of explanations needed to understand and create a trial, a few participants mentioned that they would want "more information on what a study is", explanations on "implications of changing settings" or "how the results would be calculated", and more clarity on the terms alternating and counterbalanced that are used under the advanced trial schedule settings in the app. Some suggestions were made that could inspire future versions of the app to include designing the app like a game, the ability "to do interventions together with a friend/family member" as well as recommendations about scheduling and trial settings depending on the selected intervention.

# Supplementary Figures

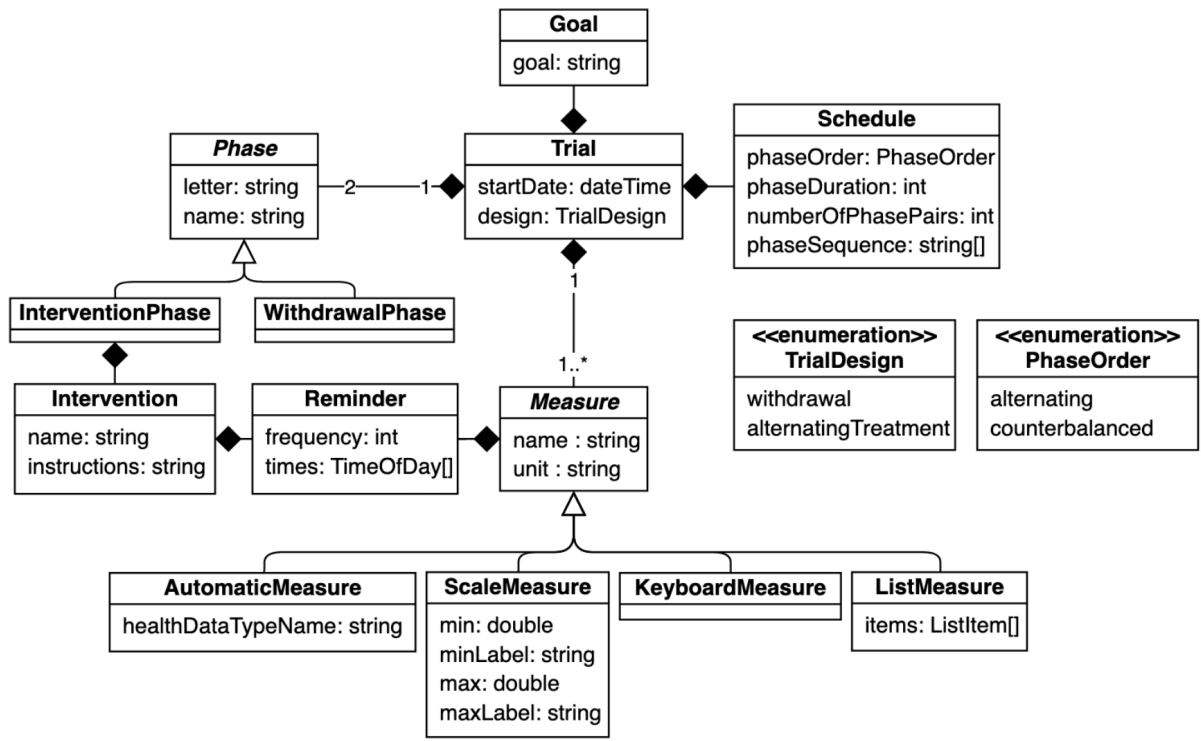

*Supplementary Figure 1. UML class diagram representing the data model of the Trial class and the classes it is composed of in StudyMe.*

The rectangles in the diagram represent classes and their properties. A line with a white triangle represents inheritance between classes, meaning the class below inherits the properties of the class above and that its objects are used in place of the upper class. A line with a black diamond at the end represents that the objects of the class on the end with a diamond are composed of objects of the class on the other end. Numbers on the lines, the multiplicities, represent how many objects are involved in the composition. One-to-one multiplicities are omitted from the diagram.

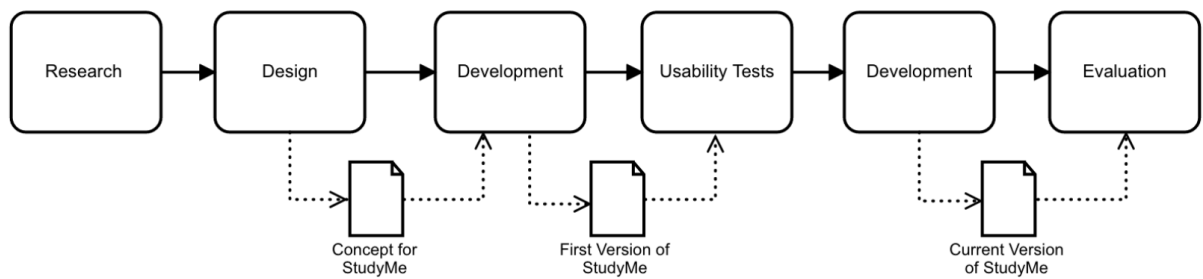

*Supplementary Figure 2. Overview of the steps of the iterative development process of StudyMe.*

# Supplementary Video

*Supplementary Video 1. Screen capture illustrating all steps in the StudyMe Health app in onboarding, generating, and running a trial.*